\documentclass[sigconf]{acmart}
\usepackage{tcolorbox}
\tcbuselibrary{listings,skins,breakable}
\usepackage{listings}
\usepackage{xcolor}

\newtcolorbox{promptbox}[1][]{
    colback=blue!5!white,
    colframe=blue!75!black,
    fonttitle=\bfseries,
    breakable,
    enhanced,
    attach boxed title to top left={yshift=-2mm, xshift=5mm},
    boxed title style={colback=blue!75!black},
    listing only,
    listing options={
        basicstyle=\ttfamily\small,
        breaklines=true,
        breakatwhitespace=true,
        columns=flexible
    },
    #1
}

\AtBeginDocument{%
  }

\setcopyright{acmlicensed}
\copyrightyear{2026}
\acmYear{2026}
\setcopyright{cc}
\setcctype{by}
\acmConference[WWW Companion '26] {Companion Proceedings of the ACM Web Conference 2026}{April 13--17, 2026}{Dubai, United Arab Emirates.}
\acmBooktitle{Companion Proceedings of the ACM Web Conference 2026 (WWW Companion '26), April 13--17, 2026, Dubai, United Arab Emirates}
\acmISBN{979-8-4007-2308-7/2026/04}
\acmDOI{10.1145/3774905.3795730}

\begin{document}

\title{Campaign-2-PT-RAG: LLM-Guided Semantic Product Type Attribution for Scalable Campaign Ranking}

\author{Yiming Che}
\authornote{equal contribution}
\affiliation{
  \institution{Walmart Global Tech}
  \city{Bentonville, AR}
  \country{USA}
}
\email{yiming.che@walmart.com}

\author{Mansi Ranjit Mane}
\authornotemark[1]
\affiliation{
  \institution{Walmart Global Tech}
  \city{Sunnyvale, CA}
  \country{USA}
}
\email{mansi.ranjit.mane@walmart.com}

\author{Keerthi Gopalakrishnan}
\authornotemark[1]
\affiliation{
  \institution{Walmart Global Tech}
  \city{Sunnyvale, CA}
  \country{USA}
}
\email{k.goppalakrishnan@walmart.com}

\author{Parisa Kaghazgaran}
\authornotemark[1]
\affiliation{
  \institution{Walmart Global Tech}
  \city{Sunnyvale, CA}
  \country{USA}
}
\email{parisa.kaghazgaran@walmart.com}

\author{Murali Mohana Krishna Dandu}
\authornotemark[1]
\affiliation{
  \institution{Walmart Global Tech}
  \city{Sunnyvale, CA}
  \country{USA}
}
\email{murali.dandu@walmart.com}

\author{Archana Venkatachalapathy}
\authornotemark[1]
\affiliation{
  \institution{Walmart Global Tech}
  \city{Sunnyvale, CA}
  \country{USA}
}
\email{Archana.Venkatachala@walmart.com}

\author{Sinduja Subramaniam}
\affiliation{
  \institution{Walmart Global Tech}
  \city{Sunnyvale, CA}
  \country{USA}
}
\email{sinduja.subramaniam@walmart.com}

\author{Yokila Arora}
\affiliation{
  \institution{Walmart Global Tech}
  \city{Sunnyvale, CA}
  \country{USA}
}
\email{yokila.arora@walmart.com}

\author{Evren Korpeoglu}
\affiliation{
  \institution{Walmart Global Tech}
  \city{Sunnyvale, CA}
  \country{USA}
}
\email{ekorpeoglu@walmart.com}

\author{Sushant Kumar}
\affiliation{
  \institution{Walmart Global Tech}
  \city{Sunnyvale, CA}
  \country{USA}
}
\email{sushant.kumar@walmart.com}

\author{Kannan Achan}
\affiliation{
  \institution{Walmart Global Tech}
  \city{Sunnyvale, CA}
  \country{USA}
}
\email{kannan.achan@walmart.com}
\renewcommand{\shortauthors}{Yiming Che et al.}

\begin{abstract}
E-commerce campaign ranking models require large-scale training labels indicating which users purchased due to campaign influence. However, generating these labels is challenging because campaigns use creative, thematic language that does not directly map to product purchases. Without clear product-level attribution, supervised learning for campaign optimization remains limited. We present \textbf{Campaign-2-PT-RAG}, a scalable label generation framework that constructs user--campaign purchase labels by inferring which product types (PTs) each campaign promotes. The framework first interprets campaign content using large language models (LLMs) to capture implicit intent, then retrieves candidate PTs through semantic search over the platform taxonomy. A structured LLM-based classifier evaluates each PT's relevance, producing a campaign-specific product coverage set. User purchases matching these PTs generate positive training labels for downstream ranking models. This approach reframes the ambiguous attribution problem into a tractable semantic alignment task, enabling scalable and consistent supervision for downstream tasks such as campaign ranking optimization in production e-commerce environments. Experiments on internal and synthetic datasets, validated against expert-annotated campaign–PT mappings, show that our LLM-assisted approach generates high-quality labels with 78--90\% precision while maintaining over 99\% recall. 
\end{abstract}




\maketitle

\section{Introduction}

E-commerce platforms increasingly rely on campaign experiences delivered through multiple user-facing channels, such as splash pages, home pages, and cart pages, to promote curated collections, seasonal promotions, and other thematic campaigns that guide product discovery and drive user conversion. Figure~\ref{fig:architecture} (left) illustrates an example of such campaigns presented as instant messages shown to users upon opening the app (i.e., a splash or pop-up page). Unlike item-level recommendations, campaigns are broad in scope: a single campaign may promote dozens of product types, often conveyed through thematic, creative, or metaphorical language rather than explicit product descriptions. This heterogeneity makes it difficult to determine whether a user has “purchased from” a campaign and, consequently, to construct reliable user–campaign labels for supervised campaign-ranking models.

In current industry practice, building such labels requires manual curation. Analysts examine a campaign’s content, determine which product types it intends to promote, and then inspect post-exposure user purchases. While feasible for a small number of curated campaigns, this approach is labor-intensive, inconsistent across annotators, slow to update, and impractical at the scale and pace of modern e-commerce environments. A key barrier to automation is the semantic alignment problem. Campaigns are written using creative or thematic language, e.g., taglines, mood statements, lifestyle imagery, which rarely map cleanly onto the structured product type (PT) taxonomy used in production systems. Embedding-based retrieval partially mitigates this gap but often over-retrieves broad concepts (e.g., ``home goods'') or misses subtle associations (e.g., ``festival essentials'' implicitly promoting portable speakers and hydration gear). Embedding models lack the ability to explicitly reason about implicit intent, contextual cues, and hierarchical relationships within the product taxonomy, often leading to either over-generalized or incomplete mappings.

To address this, we allow an LLM to interpret the campaign holistically and generate a natural-language explanation of what the campaign is promoting. This interpretation step captures latent themes and inferred product semantics that are not explicitly mentioned in the campaign text. When combined with retrieval results, it enables more accurate and interpretable mappings between campaign content and specific product types.

Building on this insight, we introduce \texttt{Campaign-2-PT-RAG}, a Retrieval-Augmented Generation (RAG) framework that infers campaign–PT alignments using semantic retrieval and structured LLM reasoning. The PT taxonomy provides canonical descriptions and hierarchical relationships that enable consistent semantic grounding. The pipeline retrieves candidate PTs based on the cosine similarity of the campaign interpretation embedding and the PT embeddings with a low threshold to include possible relevant PTs and then applies the LLM-based relevance classification to determine whether each PT is strongly relevant, weakly relevant or irrelevant. This multi-level relevance judgment mitigates retrieval noise, resolves ambiguous matches, and captures nuanced thematic associations that embedding models alone cannot. The resulting PT coverage is then used to compute user–campaign labels based on post-exposure purchases.

Experiments on both internal and synthetic datasets show that \texttt{Campaign-2-PT-RAG} substantially improves PT-mapping precision, recall, and coverage relative to traditional non-LLM-assisted baselines. These improvements translate to reliable and scalable label construction suitable for downstream campaign-ranking applications in production. Our main contributions are:
\begin{itemize}
    \item To the best of our knowledge, this work is the first to systematically study user–campaign purchase labeling in e-commerce, a problem that has been largely overlooked compared to item-level recommendation and ranking.
    \item We propose \texttt{Campaign-2-PT-RAG}, a scalable framework that combines LLM-based campaign interpretation with retrieval and structured LLM relevance reasoning over a product-type taxonomy to infer campaign-level product coverage.
    \item We demonstrate substantial improvements in PT coverage precision, recall, and semantic coherence over traditional lexical and embedding-based baselines using both real-world campaigns and synthetic evaluations
    \item Beyond label construction, Campaign-2-PT-RAG opens the door to campaign-aware learning in large-scale recommender systems by transforming unstructured campaign content into structured, interpretable product-type representations that support campaign-aware ranking, evaluation, and long-horizon optimization.
\end{itemize}

\section{Related Work}

Our work relates to previous research in three areas: (1) retrieval methods for semantic matching, (2) retrieval-augmented generation and LLM reasoning, and (3) LLM-based evaluation.

\subsection{Retrieval for Semantic Matching}

Semantic matching in information retrieval is commonly addressed through lexical or dense retrieval approaches. Classical lexical retrieval methods such as BM25~\citep{robertson2009probabilistic} rank documents based on exact term matching and remain strong baselines in many retrieval settings. Dense bi-encoder models~\citep{reimers2019sentence, karpukhin2020dense} enable efficient approximate nearest-neighbor retrieval by embedding queries and candidates into a shared vector space, allowing semantic similarity to be computed at scale.

These retrieval techniques have proven effective for search and recommendation tasks where user intent is explicitly expressed. However, they primarily optimize similarity under surface-level or distributional matching assumptions. In settings such as campaign understanding, where intent is often implicit, thematic, and multi-faceted, retrieval alone tends to over-select broadly related product types and lacks an explicit mechanism to distinguish relevance strength or reason over hierarchical relationships within a product taxonomy. Our work builds on this retrieval foundation but augments it with structured LLM-based reasoning to explicitly model relevance strength and reduce semantic ambiguity.

\subsection{Retrieval-Augmented Generation and LLM Reasoning}

RAG integrates large language models with external knowledge sources to improve grounding and reasoning~\citep{lewis2020rag, asai2024self, tang2024self}. Recent surveys provide a comprehensive overview of RAG architectures, challenges, and performance trends in modern recommendation systems~\citep{gupta2024comprehensive,gao2023retrieval}. RAG-based methods have been widely applied to knowledge-intensive tasks such as open-domain question answering and contextual generation, where retrieved context guides generation and evidence aggregation.

In e-commerce applications, LLMs have been explored for content and user intent understanding \citep{loughnane2024explicit,brinkmann2024using,zhu2020multimodal} and conversational recommendation \citep{wang2023rethinking, zhao2024recommender, liu2023conversational}. Most existing work, however, focuses on generating responses or ranking items rather than explicitly reasoning about structured relevance within a predefined taxonomy. In contrast, our approach applies RAG to infer campaign-level product-type coverage and leverages LLMs to perform explicit relevance reasoning, categorizing product types as strongly relevant, weakly relevant, or irrelevant. This formulation allows LLMs to go beyond surface-level semantic matching and address the granularity and ambiguity inherent in campaign content.

\subsection{LLM-based Evaluation}

Evaluating semantic relevance is challenging when ground truth is incomplete, subjective, or expensive to obtain. Recent studies show that large language models can serve as effective evaluators for tasks such as summarization quality, reasoning correctness, and retrieval relevance, often correlating well with human judgments \citep{li2025generation, rahmani2024judgeblender, arabzadeh2025human}. LLM-based evaluation has therefore emerged as a practical complement to traditional human annotation.

Following this line of work, we adopt an LLM-as-judge protocol to evaluate semantic alignment between campaign content and predicted product types, particularly in settings where explicit labels are unavailable. This approach provides an additional layer of semantic assessment, helping to validate model predictions beyond standard precision and recall metrics.

\section{Methodology}

We present \texttt{Campaign-2-PT-RAG}, a retrieval-augmented framework for inferring
product-type coverage from campaign content and constructing scalable user--campaign
purchase labels. The framework addresses the semantic gap between creative campaign
content and structured product taxonomies by combining LLM-based campaign
interpretation, semantic retrieval, and structured relevance reasoning. We first
formalize the problem and then describe each component of the pipeline. Finally, we provide an overview of the end-to-end system architecture, illustrating how campaign signals are ingested, processed through the LLM-assisted campaign-to-PT inference pipeline, and integrated into downstream ranking and serving components

\subsection{Problem formulation}
Let $c$ denote a campaign containing marketing content in the format \texttt{Campaign Title | Campaign Content}. Let $\mathcal{T}$ be the product-type taxonomy, where each node consists of a product category, family, and type (see Sec.~\ref{sec:knowledge_base}). The concepts in $\mathcal{T}$ are structured hierarchically, with higher-level categories encompassing broader product families and lower-level types representing fine-grained distinctions. Users generate purchase events over time. For a user $u$, let $P_u$ denote the set of purchased PTs. We aim to:
\begin{enumerate}
    \item Infer the set of PTs promoted by campaign $c$:
    \begin{equation}
        \text{PT}(c) \subseteq \mathcal{T},
    \end{equation}
    \item Construct a user-campaign label indicating whether user $u$ purchased from the inferred PTs:
    \begin{equation}
        y_{u,c} = \mathbf{1}\left(P_u \cap \text{PT}(c) \neq \emptyset \right).
    \end{equation}
\end{enumerate}

The challenge lies in mapping the content of the creative campaign to the correct PTs in $\mathcal{T}$.

\subsection{Framework Overview}

\begin{figure*}
  \includegraphics[width=\textwidth]{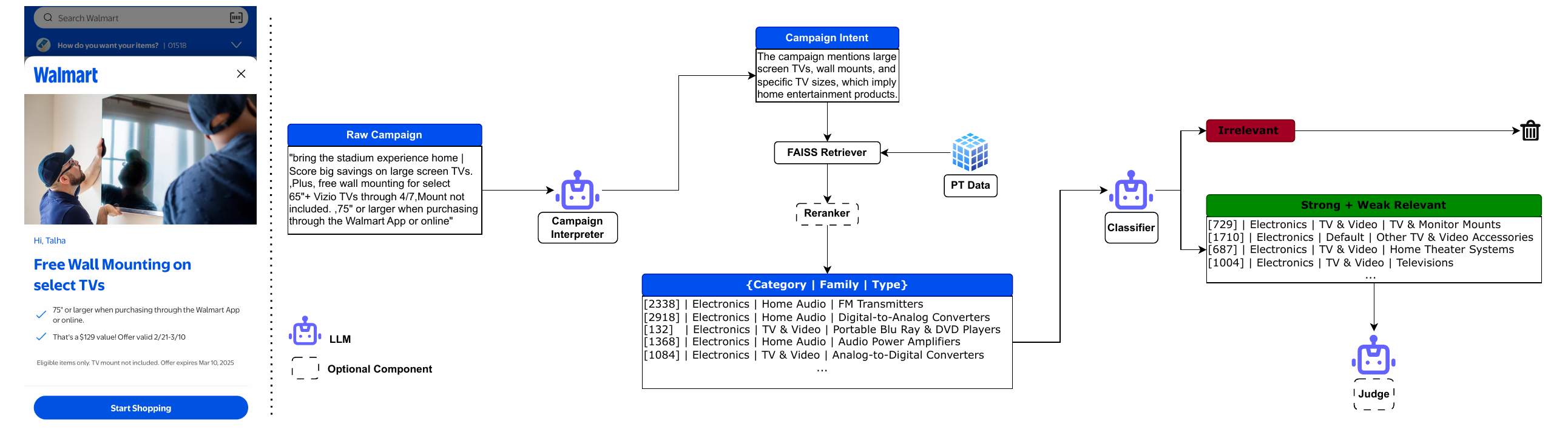}
  \caption{(Left) Illustration of a real-world e-commerce marketing campaign as presented on a splash page. (Right) Architecture of the proposed Campaign-2-PT-RAG framework for automated product curation.}
  \label{fig:architecture}
\end{figure*}

\texttt{Campaign-2-PT-RAG} comprises four sequential stages: (1) LLM-based campaign interpretation to capture explicit and implicit intent; (2) semantic retrieval of candidate PTs from the taxonomy; (3) candidates reranking; and (4) LLM-based relevance classification of retrieved PTs. Fig.~\ref{fig:architecture} (right) illustrates the overall architecture. The framework leverages the structured PT taxonomy as a knowledge base, enabling consistent semantic grounding. The LLM interpretation step enriches the retrieval query with contextual understanding, while the relevance classification step refines candidate selection through explicit reasoning. This combination addresses the granularity mismatch between campaign text and structured product representations, yielding robust and interpretable PT coverage estimates.

\subsection{Product-Type Knowledge Base}
\label{sec:knowledge_base}
The product-type taxonomy $\mathcal{T}$ serves as a structured knowledge base providing hierarchical relationships. Each PT node consists of \texttt{Category}: high-level grouping, \texttt{Family}: intermediate semantic grouping and \texttt{Type}: fine-grained product type. They are represented as concatenated text fields:
\[
\texttt{Category} \ | \ \texttt{Family} \ | \ \texttt{Type}
\]

For example, a PT node might be represented as \texttt{Electronics | Audio Equipment | Wireless Headphones}. This structured representation provides a compact yet expressive semantic backbone. Because campaigns promote concepts rather than specific stock keeping unit (SKU), PT-level alignment is both more stable across catalog updates and more robust to item-level textual noise.

In our setting, the knowledge base contains $7,147$ PTs, making manual campaign labeling extremely labor-intensive and impractical.

\subsection{Campaign-2-PT Inference via RAG}
\label{subsec:Campaign-2-PT}
Campaign content often employs thematic or creative language that does not directly reference specific PT names. To address this, we apply an LLM to interpret the campaign holistically and generate a semantic summary $s_c = \texttt{LLM\_Interpret}(c)$. The summary captures latent themes and both explicit and implicit product intents, providing a richer and more context-aware retrieval query than the raw campaign text alone.

Using the interpreted summary $s_c$, we retrieve a set of candidate PT nodes by computing cosine embedding similarity. Each PT node $t$ is embedded using its structured fields, and we retain all PTs whose cosine similarity with the campaign embedding exceeds a predefined threshold $\tau$:
\begin{equation}
  R(c) = \left\{\, t \in \mathcal{T} \;\middle|\;
  Cos\left(\text{Embed}(s_c),\ \text{Embed}(t)\right) \ge \tau \right\}.
\end{equation}

We choose $\tau$ to favor high recall, with precision improved by subsequent processing. In addition to embedding-based retrieval, we apply a cross-encoder semantic reranker \citep{nogueira2019passage} that computes pairwise relevance scores between the campaign text and PT descriptions. Cross-encoders capture fine-grained semantic interactions through joint encoding of input pairs and have been shown to be effective for neural reranking in information retrieval tasks. In our pipeline, reranking refines the candidate set and provides a more semantically ordered input for downstream LLM-based relevance classification. While reranking provides consistent but modest performance improvements (see Sec.~\ref{sec:ablation}), it incurs additional computational cost and latency. Hence, this component is treated as optional depending on deployment constraints.

For each retrieved PT $t \in R(c)$, we use an LLM to classify its relevance to campaign $c$. Given $(s_c, t)$, the LLM outputs one of:
\[
\{\texttt{strong relevance},\ \texttt{weak relevance},\ \texttt{irrelevant}\}.
\]
Formally,
\begin{equation}
    \text{Rel}(c, t) = \texttt{LLM\_Classify}(s_c, t).
\end{equation}
The LLM-based relevance classifier performs explicit semantic reasoning beyond surface-level similarity. Specifically, it infers implicit product intent expressed through creative or thematic campaign language, reasons over the hierarchical structure of the product-type taxonomy to resolve granularity mismatches, distinguishes transactional relevance from loosely related thematic associations, and incorporates negative evidence to filter product types that are semantically similar but not promoted by the campaign. This reasoning step corrects retrieval noise, resolves hierarchical ambiguity, and identifies implicit thematic relationships. We define the final PT coverage as:
\begin{equation}
    \text{PT}(c) = 
    \left\{
        t \in R(c) :
        \text{Rel}(c, t) \in \{\text{strong},\ \text{weak}\}
    \right\}.
\end{equation}
Given the inferred PT coverage, we assign a binary label indicating whether user $u$ made a relevant purchase after exposure to campaign $c$: $y_{u,c} = \mathbf{1}\left(P_u \cap \text{PT}(c) \neq \emptyset\right)$.

\subsection{System Architecture}
\label{app:sys_architecture}

\begin{figure*}
  \centering
  \includegraphics[width=\textwidth]{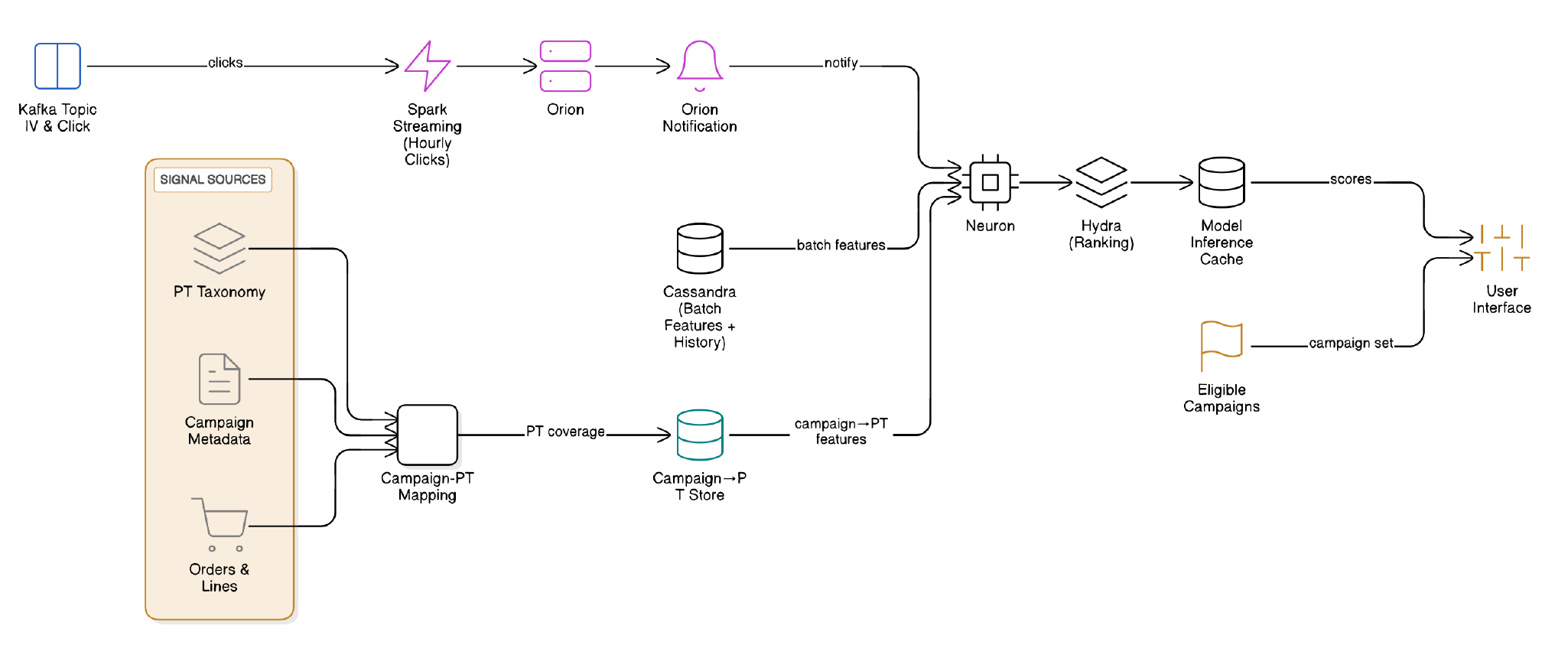}
  \caption{End-to-end system architecture integrating click signal processing with campaign-to-PT mapping. Named components (e.g., Orion, Neuron) denote logical orchestration and inference services.}
  \label{fig:sys_architecture}
\end{figure*}

Figure~\ref{fig:sys_architecture} presents the end-to-end system architecture
integrating real-time user interaction signals with the \texttt{Campaign-2-PT-RAG} pipeline for scalable campaign attribution and ranking. The overall workflow can be divided into three stages: signal ingestion and triggering, campaign-to-PT mapping, and decisioning and serving.

\paragraph{Signal Ingestion and Triggering.}
User impression and click events are ingested through a unified Kafka stream.
Click signals are processed via an hourly Spark job, while impression signals are forwarded directly to the orchestration layer to minimize latency. Both signals are consolidated in \textit{Orion}, an internal event orchestration service that coordinates downstream processing. \textit{Orion}, a real-time orchestration and triggering service, that emits notifications that trigger inference in \textit{Neuron} and \textit{Hydra}, which are the campaign inference and decisioning service, which also consumes batch features and historical aggregates stored in \textit{Cassandra}, a distributed key–value store.

\paragraph{Campaign-to-PT Mapping via LLM-RAG.}
Campaign-to-PT mapping is performed using the proposed \texttt{Campaign-2-PT-RAG}
pipeline described in Section~\ref{subsec:Campaign-2-PT}. For each incoming campaign, the LLM first interprets campaign metadata to infer implicit intent, retrieves candidate product types from the taxonomy, and applies structured relevance classification to assign strong, weak, or irrelevant labels. The resulting campaign-level PT coverage provides a compact and interpretable representation of campaign intent, which serves as a key semantic signal for downstream task.

\paragraph{Decisioning and Serving.}
\textit{Neuron} combines real-time triggers, batch features, and campaign–PT signals to produce candidate scores, which are ranked by Hydra and cached for low-latency serving through the personalization layer under campaign eligibility constraints.

While we refer to internal system names for concreteness, these components correspond to standard functions in large-scale recommender architectures, including event orchestration, model inference, and ranking, and can be implemented using equivalent services in other production environments.

\section{Experiments}

We evaluate \texttt{Campaign-2-PT-RAG} on the task of campaign-PT mapping, focusing on the intrinsic quality of inferred PT sets. Our evaluation framework consists of three components: (1) \textbf{Human-labeled campaigns}: measures accuracy against curated ground truth, (2) \textbf{Synthetic campaigns}: measures recovery of implicit PT associations, and (3) \textbf{LLM-as-judge evaluation}: provides semantic relevance judgments to complement incomplete human labels. Finally, the ablation studies are detailed in the Section~\ref{sec:ablation}. 

\subsection{Experiment Setup}

We use GPT-4o (API version 2024--10--21) as the LLM for campaign interpretation and relevance classification. For neural reranking, we employ the \texttt{cross-encoder/ms-marco-MiniLM-L6-v2} model from Hugging Face, a lightweight cross-encoder trained on the MS MARCO passage ranking task. The embedding model used for retrieval is \texttt{sentence-transformers/all-mpnet-base-v2}, a widely used bi-encoder model for semantic search. The PT knowledge base includes $7,147$ internal PTs.

We compare \texttt{Campaign-2-PT-RAG} against the following baselines:

\begin{itemize}
    \item \textbf{BM25 Retrieval}: A lexical retrieval baseline that ranks product types
    using BM25 over PT text fields. This method serves as a strong keyword-based
    baseline.
    \item \textbf{LLM Zero-Shot Classification}: The LLM is provided with the full
    product-type list and asked to directly select relevant PTs for a campaign.
    This approach is susceptible to position bias and long-context degradation.
    \item \textbf{Embedding Retrieval}: Product types are directly retrieved using cosine
    similarity between embeddings of raw campaign and PT embeddings, with multiple
    similarity thresholds.
\end{itemize}
All LLM components are used in a zero-shot or prompt-based manner, without
fine-tuning on campaign or purchase data.

\subsection{Evaluation for Human-Labeled Campaigns}

Let $\text{PT}_{\text{true}}(c)$ denote the ground truth PT set for campaign $c$ from human labeling and $\text{PT}_{\text{pred}}(c)$ the predicted set, i.e., the union of strongly relevant set and weakly relevant set from LLM classification. We report the following metrics.

\paragraph{Precision}
\begin{equation}
\text{Precision}(c) =
\frac{
|\text{PT}_{\text{pred}}(c) \cap \text{PT}_{\text{true}}(c)|
}{
|\text{PT}_{\text{pred}}(c)|
}.
\end{equation}

\paragraph{Recall}
\begin{equation}
\text{Recall}(c) =
\frac{
|\text{PT}_{\text{pred}}(c) \cap \text{PT}_{\text{true}}(c)|
}{
|\text{PT}_{\text{true}}(c)|
}.
\end{equation}

\paragraph{F1 Score}
\begin{equation}
\text{F1}(c) =
2 \cdot 
\frac{\text{Precision}(c)\cdot \text{Recall}(c)}
{\text{Precision}(c) + \text{Recall}(c)}.
\end{equation}

\paragraph{Semantic Coherence}
It measures whether the predicted PT set is semantically consistent:
\begin{equation}
\text{Coherence}(c) =
\frac{1}{|S|}
\sum_{(t_1,t_2)\in S}
\text{cosine}(t_1,t_2),
\end{equation}
where $S$ is the set of unordered PT embedding pairs in $\text{PT}_{\text{pred}}(c)$. Semantic coherence complements precision/recall by assessing thematic consistency even when ground truth is incomplete.

\subsection{Evaluation for LLM-as-Judge }
\label{sec:LLM-as-Judge}

While human-labeled PT sets provide an essential source of ground truth, they are inherently incomplete: annotators may overlook weakly relevant PTs, disagree on implicit product associations, or miss long-tail categories that are semantically aligned with the campaign. To complement these limitations, we adopt an \textit{LLM-as-judge} protocol to provide an additional semantic evaluation of model predictions. Given a campaign $c$ and predicted PT $t$, an LLM assigns: strongly relevant, weakly relevant and irrelevant. This produces a relevance label $\texttt{LLMJudge}(c,t)$ that serves as an external semantic assessment. We compute the following metrics:

\paragraph{LLM-Precision.}
The fraction of predicted PTs judged as strongly or weakly relevant, i.e., 
\begin{equation}
\text{LLM-Precision}(c) =
\frac{
|\{ t : \texttt{LLMJudge}(c,t) \neq \text{irrele.} \land t \in \text{PT}_{\text{pred}}(c) \}|
}{
|\text{PT}_{\text{pred}}(c)|}
\end{equation}

\paragraph{LLM-Recall.}
The proportion of LLM-relevant PTs recovered by the model, computed over 
the union of PTs predicted by all systems, i.e., 
\begin{equation}
\text{LLM-Recall}(c) =
\frac{
|\{ t : \text{LLMJudge}(c,t) \neq \text{irrele.} \land t \in \text{PT}_{\text{pred}}(c) \}|
}{
|\{ t : \text{LLMJudge}(c,t) \neq \text{irrele.}\}|
}.
\end{equation}

\paragraph{LLM Score.}
A holistic 0--1 rating of how well the predicted PT set matches campaign intent.

\subsection{Walmart Internal Campaign Evaluation}
\label{sec:internal campaigns}

We evaluate \texttt{Campaign-2-PT-RAG} on a set of real-world Walmart campaigns manually annotated by domain experts at Walmart. In total, we sample eight campaigns covering diverse Walmart themes such as Electronics, Fresh produce, Home Outdoor, Auto-Care, Back-to-School preparation, and Home Essentials.

\textbf{Fresh produce} \textit{Groceries guaranteed fresh or your money back | Fresh produce, delivered daily to our stores.}



Each campaign is independently annotated by at least two experts, who identify product types relevant to the campaign intent. The final ground-truth PT set is constructed by aggregating the relevant PTs identified by all annotators and resolving disagreements through discussion. Inter-annotator agreement is measured using Jaccard distance ($0.9412 \pm 0.1326$). 

Table~\ref{tab:human-eval} summarizes precision, recall, and F1 scores, reported as mean $\pm$ standard deviation across campaigns, using human annotations as ground truth. Retrieval-only baselines achieve high recall but relatively low precision, indicating substantial over-selection of loosely related product types. For example, in the fresh produce campaign (see Section~\ref{sec:internal campaigns}), which is interpreted by the LLM as \textit{focusing on groceries, including fresh produce and daily essentials}, retrieval-based methods using cosine similarity thresholds of $0.5$ and $0.3$ incorrectly
include product types such as snack boxes and emergency food. Although these categories are broadly food-related, they are not aligned with the specific intent of the campaign. In addition, some clearly unrelated product types, such as \texttt{Office \& Stationery | Money Handling | Cash Registers}, appear close to the raw campaign embedding in the vector space due to superficial lexical or distributional similarity, yet are semantically distant from the campaign intent. In contrast, \texttt{Campaign-2-PT-RAG} effectively filters out both
loosely related and spurious matches through structured LLM-based relevance reasoning, resulting in substantially improved precision while preserving high recall. We also observe that using a high retrieval threshold (e.g., $0.7$) yields very small predicted PT sets, which results in coincident precision, recall, and F1 values.

Lexical retrieval method BM25 is particularly prone to spurious matches. In the same fresh produce campaign, BM25 retrieves product types like \texttt{Office \& Stationery | Money Handling | Money Deposit Bags} solely due to lexical overlap with the term ``money'' despite being entirely unrelated to food or grocery content. In contrast, \texttt{Campaign-2-PT-RAG} successfully filters out these false positives through LLM-based relevance reasoning, resulting in substantially improved precision while preserving recall.

The zero-shot LLM baseline performs poorly across all metrics, reflecting the
difficulty of directly reasoning over thousands of product types without
retrieval or structured context. When presented with long, unfiltered PT lists,
the LLM exhibits degraded precision due to limited context capacity and a lack
of explicit mechanisms to compare relevance across many candidates. In addition,
this approach incurs substantially higher token usage, as the entire PT list
must be included in the prompt for each campaign, making it computationally
inefficient.

To further assess semantic quality beyond exact overlap with human labels, we apply the LLM-as-judge evaluation described in Section~\ref{sec:LLM-as-Judge}.  We first measure the agreement between human annotations and LLM-judge annotations across eight Walmart internal campaigns using Jaccard similarity. The resulting agreement score of $0.9376 \pm 0.1120$ indicates strong consistency between the LLM judge and human annotators, suggesting that the LLM reliably captures campaign-level product-type relevance in this setting.

The results in terms of LLM-as-judge are shown in Table~2. \texttt{Campaign-2-PT-RAG} achieves the best results in all metrics, indicating that the predicted PT sets are not only accurate with respect to human annotations but also semantically consistent and well-aligned with campaign intent. These results highlight the benefit of combining semantic retrieval with structured LLM reasoning for campaign-level product type inference in real-world settings.


\begin{table}
\caption{PT-mapping quality on real Walmart campaigns with human annotations. For retrieval-based models, the value in parentheses denotes the cosine similarity threshold used for candidate selection. The best performance for each metric is shown in bold.}
  \label{tab:human-eval}
  \setlength{\tabcolsep}{1pt}
  \begin{tabular}{l ccc}
    \toprule
    Model & Precision & Recall & F1\\
    \midrule
    BM25                    & $0.2013 \pm 0.1246$ & $0.9532 \pm 0.0120$ & $0.3276 \pm 0.1293$ \\
    Zero-Shot LLM           & $0.0415 \pm 0.0312$ & $0.5043 \pm 0.4992$ & $0.0932 \pm 0.0913$ \\
    Retrieval $(0.3)$      & $0.4015 \pm 0.2829$ & \boldmath{$0.9973 \pm 0.0073$} & $0.5263 \pm 0.2640$ \\
    Retrieval $(0.5)$      & $0.4020 \pm 0.2523$ & $0.9922 \pm 0.0117$ & $0.5328 \pm 0.2421$ \\
    Retrieval $(0.7)$      & $0.2500 \pm 0.4629$ & $0.2500 \pm 0.4629$ & $0.2500 \pm 0.4629$ \\
    \textbf{Campaign-2-PT}  & \boldmath{$0.8934 \pm 0.0825$} & $0.9756 \pm 0.0172$ & \boldmath{$0.9412 \pm 0.0376$} \\
    \bottomrule
  \end{tabular}
\end{table}

\begin{table*}
\caption{PT-mapping quality on real Walmart campaigns with LLM judge. For retrieval-based models, the value in parentheses denotes the cosine similarity threshold used for candidate selection. The best performance for each metric is shown in bold.}
\label{tab:LLM-eval}
\centering
  \begin{tabular}{lccccc}
    \toprule
    Model & Precision & Recall & F1 & Coherence &LLM Score\\
    \midrule
    BM25  & $0.2100 \pm 0.0946$ & $1.0000 \pm 0.0000$ & $0.3386 \pm 0.1285$ & -- & $0.5714 \pm 0.1955$ \\
    Zero-Shot LLM & $0.0525 \pm 0.0514$ & $0.5153 \pm 0.4896$ & $0.0949 \pm 0.0919$ & -- & $0.1500 \pm 0.1309$ \\
    Retrieval ($0.3$) & $0.4075 \pm 0.2914$ & $0.9972 \pm 0.0077$ & $0.5284 \pm 0.2740$ & $0.6730 \pm 0.0654$ & $0.6500 \pm 0.2000$ \\
    Retrieval ($0.5$) & $0.4041 \pm 0.2543$ & $0.9942 \pm 0.0119$ & $0.5359 \pm 0.2466$ & $0.6743 \pm 0.0644$ & $0.7250 \pm 0.1982$ \\
    Retrieval ($0.7$) & $0.2500 \pm 0.4629$ & $0.2500 \pm 0.4629$ & $0.2500 \pm 0.4629$ & $0.2253 \pm 0.1355$ & $0.2125 \pm 0.4016$ \\
    \textbf{Campaign-2-PT} & \boldmath{$0.9054 \pm 0.0600$} & \boldmath{$1.0000 \pm 0.0000$} & \boldmath{$0.9495 \pm 0.0326$} & \boldmath{$0.7976 \pm 0.0332$} & \boldmath{$0.8900 \pm 0.0566$}  \\
    \bottomrule
  \end{tabular}
\end{table*}

\subsection{Synthetic Campaign Evaluation}

In addition to human-labeled real-world campaigns, we evaluate \texttt{Campaign-2-PT-RAG} using synthetically generated campaign descriptions designed to resemble real-world e-commerce marketing. These campaigns are generated by ChatGPT 5.2 in a concise, transactional style (e.g., campaign title and brief description), similar to Walmart campaigns used in production systems, e.g., 

\textit{Small-Space Living Solutions Essentials | Entertain easily with casual pieces and accessories that are ready when guests arrive, with free delivery available.}

\textit{On-the-Go Essentials | Snack healthier with convenient options you can grab on busy days without sacrificing taste Easy returns available.}

\textit{Travel Light \& Ready | Pack lighter for weekend trips with travel-ready items that save space and simplify transitions. Enjoy simple, reliable choices for every day.}

The synthetic campaigns are not constructed with explicit product-type labels, and therefore do not provide direct ground truth for PT relevance.

Because the promoted product domains are implicit and not directly observable
from the generation process, we rely on the LLM-as-judge protocol described in
Section~3.3 to evaluate PT-mapping quality. After each method predicts a set of
relevant product types for a campaign, an independent LLM assesses the semantic
alignment between the campaign content and the predicted PTs, producing
relevance judgments and a set-level quality score. Given the high agreement
between LLM judge assessments and human annotations observed on real-world
campaigns, we consider the LLM judge to be a reliable proxy for semantic
evaluation in this setting.

Table~3 summarizes the results of synthetic campaigns in terms of LLM-as-judge. Lexical and embedding-only baselines again exhibit high recall but lower precision and semantic coherence, reflecting their tendency to retrieve broadly related PTs under ambiguous campaign language. The zero-shot LLM baseline performs poorly, underscoring the difficulty of directly reasoning over PTs without retrieval or structured filtering. \texttt{Campaign-2-PT-RAG} consistently achieves the highest F1 score, semantic coherence, and LLM score, demonstrating improved robustness to implicit, abstract, and multi-intent campaign language. These results suggest that LLM-assisted campaign interpretation and relevance reasoning are critical for effective PT inference when explicit supervision is unavailable.

\begin{table*}
\caption{PT-mapping quality on synthetic campaigns with LLM judge. For retrieval-based models, the value in parentheses denotes the cosine similarity threshold used for candidate selection. The best performance for each metric is shown in bold.}
\label{tab:synthetic-eval}
  \begin{tabular}{lccccc}
    \toprule
    Model & Precision & Recall & F1 & Coherence & LLM Score \\
    \midrule
    BM25  & $0.3442 \pm 0.2127$ & $0.9837 \pm 0.1050$ & $0.4738 \pm 0.2364$ & -- & $0.5869 \pm 0.1868$ \\
    Zero-Shot LLM & $0.0568 \pm 0.0496$  & $0.4416 \pm 0.3623$ & $0.0990 \pm 0.0844$ & -- & $0.3147 \pm 0.1338$  \\
    Retrieval ($0.3$) & $0.3632 \pm 0.1691$ & $0.9893 \pm 0.1001$ & $0.5106 \pm 0.1825$ & $0.7001 \pm 0.0386$ & $0.6693 \pm 0.1373$ \\
    Retrieval ($0.5$) & $0.3536 \pm 0.1680$ & $0.9953 \pm 0.0308$ & $0.5002 \pm 0.1790$ & $0.7009 \pm 0.0385$ & $0.6522 \pm 0.1395$ \\
    Retrieval ($0.7$) & $0.2081 \pm 0.3908$ & $0.2300 \pm 0.4230$ & $0.2163 \pm 0.4012$ & $0.1238 \pm 0.2975$ & $0.1940 \pm 0.3668$ \\
    \textbf{Campaign-2-PT} & \boldmath{$0.8219 \pm 0.1614$} & \boldmath{$0.9926 \pm 0.0032$} & \boldmath{$0.8718 \pm 0.1003$} & \boldmath{$0.7876 \pm 0.0335$} & \boldmath{$0.8250 \pm 0.0894$} \\
  \bottomrule
  \end{tabular}
\end{table*}

\subsection{Ablation Studies}
\label{sec:ablation}
We conduct ablation studies to quantify the contribution of each component in \texttt{Campaign-2-PT-RAG}, with results summarized in Table~\ref{tab:ablation-eval} using the LLM-as-judge evaluation on synthetic campaigns. A retrieval-only baseline achieves high recall but low precision and semantic coherence, indicating substantial over-selection of loosely related product types and confirming that dense retrieval alone is insufficient for disambiguating campaign intent. Adding LLM-based campaign interpretation before retrieval significantly improves precision and coherence by producing a semantically enriched query that better captures implicit campaign intent, though some weakly related PTs remain without explicit filtering. Introducing LLM-based relevance classification yields the largest gains in precision, F1, and LLM score, demonstrating the importance of structured reasoning over retrieved candidates to distinguish strong relevance from weak or irrelevant associations. The full \texttt{Campaign-2-PT-RAG} pipeline, which combines campaign interpretation, retrieval, reranking, and LLM-based relevance classification, achieves the best performance across all metrics, confirming that these components provide complementary benefits and that explicit campaign interpretation and LLM reasoning are the primary drivers of improved semantic alignment.

\begin{table*}
\centering
\caption{Ablation studies on PT-mapping quality on synthetic campaigns with LLM judge. For retrieval-based models, the value in parentheses denotes the cosine similarity threshold used for candidate selection. The best performance for each metric is shown in bold.}
\label{tab:ablation-eval}
  \begin{tabular}{lccccc}
    \toprule
    Model & Precision & Recall & F1 & Coherence & LLM Score \\
    \midrule
    Retrieval Only ($0.5$) & $0.3536 \pm 0.1680$ & $0.9953 \pm 0.0308$ & $0.5002 \pm 0.1790$ & $0.7009 \pm 0.0385$ & $0.6522 \pm 0.1395$ \\
    Des. + Retrieval & $0.4690 \pm 0.1848$ & $0.9972 \pm 0.0083$ & $0.6163 \pm 0.1746$ & $0.7294 \pm 0.0381$ & $0.7533 \pm 0.1124$ \\
    Des. + Retrieval + LLM & $0.7708 \pm 0.1516$ & \boldmath{$0.9996 \pm 0.0027$} & $0.8656 \pm 0.1041$ & \boldmath{$0.7607 \pm 0.0463$} & $0.8286 \pm 0.0847$ \\
    Des. + Retrieval + Reranker + LLM & \boldmath{$0.7819 \pm 0.1814$} & $0.9976 \pm 0.0022$ & \boldmath{$0.8688 \pm 0.1013$} & $0.7576 \pm 0.0435$ & \boldmath{$0.8350 \pm 0.0896$} \\
  \bottomrule
  \end{tabular}
\end{table*}

\section{Discussion and Limitations}
While \texttt{Campaign-2-PT-RAG} improves campaign–PT mapping, it relies on
LLMs whose performance and cost may vary across deployments.
Although LLM-based relevance reasoning reduces semantic ambiguity, relevance
judgments remain inherently subjective, even among human annotators. In addition,
the optional reranking and LLM inference stages introduce latency that may not be
suitable for all real-time settings. Future work will explore lighter-weight
reasoning models and tighter integration with downstream ranking objectives.

\section{Conclusion}
We introduced \texttt{Campaign-2-PT-RAG}, a RAG framework for scalable construction of user–campaign purchase labels through semantic alignment of campaign content with product-type taxonomies. By combining LLM interpretation, semantic retrieval, and structured relevance classification, the framework significantly improves the quality of campaign–PT mappings. By enabling scalable and interpretable campaign labeling, \texttt{Campaign-2-PT-RAG} provides a practical foundation for deploying user-level campaign ranking in large-scale e-commerce systems.

\bibliographystyle{ACM-Reference-Format}
\bibliography{ref}

@String{Springer = "Springer-Verlag" }

@article{robertson2009probabilistic,
  title={The Probabilistic Relevance Framework: BM25 and Beyond},
  author={Robertson, Stephen and Zaragoza, Hugo},
  journal={Foundations and Trends in Information Retrieval},
  year={2009},
  note={classic BM25 reference}
}

@article{nogueira2019passage,
  title={Passage Re-ranking with BERT},
  author={Nogueira, Rodrigo and Cho, Kyunghyun},
  journal={arXiv:1901.04085},
  year={2019}
}

@inproceedings{lewis2020rag,
  title={Retrieval-Augmented Generation for Knowledge-Intensive NLP Tasks},
  author={Lewis, Patrick and others},
  booktitle={NeurIPS},
  year={2020}
}

@inproceedings{reimers2019sentence,
  title={Sentence-BERT: Sentence Embeddings using Siamese BERT-Networks},
  author={Reimers, Nils and Gurevych, Iryna},
  booktitle={Proceedings of the 2019 Conference on Empirical Methods in Natural Language Processing and the 9th International Joint Conference on Natural Language Processing (EMNLP-IJCNLP)},
  pages={3982--3992},
  year={2019}
}

@inproceedings{karpukhin2020dense,
  title={Dense Passage Retrieval for Open-Domain Question Answering.},
  author={Karpukhin, Vladimir and Oguz, Barlas and Min, Sewon and Lewis, Patrick SH and Wu, Ledell and Edunov, Sergey and Chen, Danqi and Yih, Wen-tau},
  booktitle={EMNLP (1)},
  pages={6769--6781},
  year={2020}
}

@article{gupta2024comprehensive,
  title={A comprehensive survey of retrieval-augmented generation (rag): Evolution, current landscape and future directions},
  author={Gupta, Shailja and Ranjan, Rajesh and Singh, Surya Narayan},
  journal={arXiv preprint arXiv:2410.12837},
  year={2024}
}

@article{gao2023retrieval,
  title={Retrieval-augmented generation for large language models: A survey},
  author={Gao, Yunfan and Xiong, Yun and Gao, Xinyu and Jia, Kangxiang and Pan, Jinliu and Bi, Yuxi and Dai, Yixin and Sun, Jiawei and Wang, Haofen and Wang, Haofen},
  journal={arXiv preprint arXiv:2312.10997},
  volume={2},
  number={1},
  year={2023}
}

@article{zhao2024recommender,
  title={Recommender systems in the era of large language models (llms)},
  author={Zhao, Zihuai and Fan, Wenqi and Li, Jiatong and Liu, Yunqing and Mei, Xiaowei and Wang, Yiqi and Wen, Zhen and Wang, Fei and Zhao, Xiangyu and Tang, Jiliang and others},
  journal={IEEE Transactions on Knowledge and Data Engineering},
  volume={36},
  number={11},
  pages={6889--6907},
  year={2024},
  publisher={IEEE}
}

@inproceedings{li2025generation,
  title={From generation to judgment: Opportunities and challenges of llm-as-a-judge},
  author={Li, Dawei and Jiang, Bohan and Huang, Liangjie and Beigi, Alimohammad and Zhao, Chengshuai and Tan, Zhen and Bhattacharjee, Amrita and Jiang, Yuxuan and Chen, Canyu and Wu, Tianhao and others},
  booktitle={Proceedings of the 2025 Conference on Empirical Methods in Natural Language Processing},
  pages={2757--2791},
  year={2025}
}

@inproceedings{loughnane2024explicit,
  title={Explicit Attribute Extraction in E-Commerce Search},
  author={Loughnane, Robyn and Liu, Jiaxin and Chen, Zhilin and Wang, Zhiqi and Giroux, Joseph and Du, Tianchuan and Schroeder, Benjamin and Sun, Weiyi},
  booktitle={Proceedings of the Seventh Workshop on e-Commerce and NLP@ LREC-COLING 2024},
  pages={125--135},
  year={2024}
}

@inproceedings{brinkmann2024using,
  title={Using LLMs for the extraction and normalization of product attribute values},
  author={Brinkmann, Alexander and Baumann, Nick and Bizer, Christian},
  booktitle={European Conference on Advances in Databases and Information Systems},
  pages={217--230},
  year={2024},
  organization={Springer}
}

@inproceedings{zhu2020multimodal,
  title={Multimodal Joint Attribute Prediction and Value Extraction for E-commerce Product},
  author={Zhu, Tiangang and Wang, Yue and Li, Haoran and Wu, Youzheng and He, Xiaodong and Zhou, Bowen},
  booktitle={Proceedings of the 2020 Conference on Empirical Methods in Natural Language Processing (EMNLP)},
  pages={2129--2139},
  year={2020}
}

@inproceedings{wang2023rethinking,
  title={Rethinking the Evaluation for Conversational Recommendation in the Era of Large Language Models},
  author={Wang, Xiaolei and Tang, Xinyu and Zhao, Wayne Xin and Wang, Jingyuan and Wen, Ji-Rong},
  booktitle={Proceedings of the 2023 Conference on Empirical Methods in Natural Language Processing},
  pages={10052--10065},
  year={2023}
}

@inproceedings{liu2023conversational,
  title={Conversational recommender system and large language model are made for each other in E-commerce pre-sales dialogue},
  author={Liu, Yuanxing and Zhang, Weinan and Chen, Yifan and Zhang, Yuchi and Bai, Haopeng and Feng, Fan and Cui, Hengbin and Li, Yongbin and Che, Wanxiang},
  booktitle={Findings of the Association for Computational Linguistics: EMNLP 2023},
  pages={9587--9605},
  year={2023}
}

@inproceedings{asai2024self,
  title={Self-RAG: Learning to Retrieve, Generate, and Critique through Self-Reflection},
  author={Asai, Akari and Wu, Zeqiu and Wang, Yizhong and Sil, Avi and Hajishirzi, Hannaneh},
  booktitle={International Conference on Learning Representations},
  year={2024}
}

@article{tang2024self,
  title={Self-retrieval: End-to-end information retrieval with one large language model},
  author={Tang, Qiaoyu and Chen, Jiawei and Li, Zhuoqun and Yu, Bowen and Lu, Yaojie and Yu, Haiyang and Lin, Hongyu and Huang, Fei and He, Ben and Han, Xianpei and others},
  journal={Advances in Neural Information Processing Systems},
  volume={37},
  pages={63510--63533},
  year={2024}
}

@article{rahmani2024judgeblender,
  title={JudgeBlender: Ensembling Judgments for Automatic Relevance Assessment},
  author={Rahmani, Hossein A and Yilmaz, Emine and Craswell, Nick and Mitra, Bhaskar},
  journal={CoRR},
  year={2024}
}

@inproceedings{arabzadeh2025human,
  title={A human-ai comparative analysis of prompt sensitivity in llm-based relevance judgment},
  author={Arabzadeh, Negar and Clarke, Charles LA},
  booktitle={Proceedings of the 48th International ACM SIGIR Conference on Research and Development in Information Retrieval},
  pages={2784--2788},
  year={2025}
}

\end{document}